\documentclass[article,11pt,oneside,a4paper,]{revtex4}
\usepackage{amsfonts}
\usepackage{graphicx}
\usepackage{times}
\usepackage{color}
\usepackage{graphics}
\usepackage{amssymb}
\usepackage[brazilian]{babel}
\usepackage{slashed}

\newcommand{\be}{\begin{equation}}
\newcommand{\ee}{\end{equation}}
\newcommand{\ben}{\begin{eqnarray}}
\newcommand{\een}{\end{eqnarray}}

\begin{document}
\begin{center}
{\Large{\textbf{The bosonic and fermionic propagators under an external electromagnetic field in a Euclidean manifold}}
}
\end{center}

\author{E. B. S. Corr\^{e}a{\footnote {emersoncbpf@gmail.com}}}
\affiliation{Instituto de Ci\^{e}ncias Exatas, Universidade Federal do Sul e Sudeste do Par\'a-UNIFESSPA, 68505080, Marab\'a, PA, Brazil}
\author{C. A. Bahia{\footnote {andre.bahia@unifesspa.edu.br}}}
\affiliation{Faculdade de Engenharia elétrica, Universidade Federal do Sul e Sudeste do Par\'a-UNIFESSPA, 68505080, Marab\'a, PA, Brazil}
\author{J. A. Lourenço{\footnote {quantumlourenco@gmail.com}}}
\affiliation{Departamento de Ci\^encias Naturais, Universidade Federal do Esp\'\i rito Santo, 29932-540 Campus S\~ao Mateus, ES, Brazil}
\begin{abstract}
In this paper, we will calculate the bosonic as well as fermionic propagators under classical homogeneous and constant magnetic and electric fields in a Euclidean space. For this, we will reassess the Ritus' method for calculating the Feynman propagator.\\

\end{abstract}
\maketitle

\section{Introduction}

   It is well known that the Feynman propagator is an amount very important in Quantum Field Theory (QFT) and that, therefore it must be calculated for the most diverse situations in which quantum fields are subject, for example in periodic boundary conditions, antiperiodic boundary conditions and in external fields.
   
   Taking into account external electromagnetic fields, the propagator can be calculated using the technique developend by J. Schwinger in $1951$, the proper time method \cite{Schwinger}. The proper time method is a powerful way to calculate the propagator under external fields, but it uses one dimension additional to get the propagator: the proper time $S$ in the range $[0,+\infty]$. Thus, the expressions to the propagators are given by integrals on proper time $S$. Thise integrals can be very difficult to solve ~\cite{Schwinger,Farina}. In the decade of $1970$, V. Ritus addressed the problem of calculating the fermionic field propagator subject to an external electromagnetic field of a rather innovative and simplistic way \cite{Ritus,Ritus1,Ritus2}, namely: by the diagonalization of the Dirac operator. In other words, the method consists in to find eigenfunctions of Dirac's operator such that the propagator is written as in free form. That is, it is found a kind of Fourier transform for the operator ${\slashed{D}}$, being $D_{\mu}$ the covariant derivative. A few years ago, the Ritus' method also has been used to calculate the propagator of a charged particle of spin 1 in the context of electroweak theory \cite{Elizalde, Elizalde1} and of spin $1/2$ in low dimensions~\cite{Mexicanos}. Here, we will rescue Ritus' interesting idea for calculating Feynman propagator under external, constant and homogeneous electric and magnetic fields in the $ z $ direction. 	
   	 
   The paper is organized as follows: in section II, we present the Ritus' method. As we want to be didactic, let us
   divide it into four subsections. In subsection {\bf{A}} we calculate the free scalar propagator using the eigenfunctions of $\partial_{\mu}$ operator: the plane waves. In the subsection {\bf{B}} we calculate the scalar propagator under an external magnetic and electric classical fields along to $z$ direction. Let us calculate the eigenfunctions of operator ${D}^{2}$ and show that they form a complete set of solutions. From this set, we will be able to expand the propagator in the coordinate as well as momenta spaces. In subsection {\bf{C}}, we calculate the Feynman propagator of the free spinor field. Again, to be pedagogical, we uses the Ritus' method in the $\slashed{\partial}$ operator context, i.e., we write the free fermionic propagator by expansion over plane waves. In subsection {\bf{D}}, we calculate the spinor propagator field under an electric and magnetic external background fields in the $z$ direction, through the eigenfunctions associated to $\slashed{D}$ operator. We will make our final considerations in section III. Throughout the text, we will use a four-dimensional Euclidean space and the natural system of units in which $c=\hbar=1$. Our definitions agree with Ref.~\cite{Ramond}.
   
\section{Ritus' method}

The Ritus' method can be describe in few words as follows. The field propagator satisfy an differential equation. This differential equation has a specific operator. We solved the eigenvalue equation for this specific operator. Then, we expand the propagator in terms of   operator's eigenfunctions. To facilitate understanding, let us show how the Ritus' method works in four cases: free scalar field; scalar field under external fields $\bf{E}$ and $\bf{B}$; free spinor field; and spinor field under external fields $\bf{E}$ and $\bf{B}$.

\subsection{Free scalar field propagator}

The free Klein-Gordon equation in four-dimensional Euclidean space is given by

\begin{eqnarray*}
(-\partial_{\mu}\partial_{\mu} + m^{2} )\phi(u) = 0,
\label{Eq. K-G semcampo}
\end{eqnarray*}
where $\partial_{\mu} = (\partial_{\tau},\partial_{x},\partial_{y},\partial_{z})$  and $u = (\tau,x,y,z)$. The Feynman propagator associated to free scalar field satisfy~\cite{Ramond} 
\begin{eqnarray}
(\partial_{\mu}\partial_{\mu} - m^{2})G(u-u^{\prime}) = -\delta^{4}(u-u^{\prime}).
\label{GlivreEqDif}
\end{eqnarray}
Note that the plane wave in Euclidean space, $ \exp{(i{k}_{\alpha} u_{\alpha})} $, are eigenfunction of $\partial_{\nu}$ with $i{k}_{\nu}$ eigenvalue. Furthermore, we have $[\partial_{\mu}\partial_{\mu},\partial_{\nu}]=0$. For this, plane waves are also eigenfunctions of $\partial_{\mu}\partial_{\mu}$ operator, with $-(k_{\mu})^2$ eigenvalue. Since plane waves form a complete set of functions, we can write
\begin{eqnarray}
G(u-u^{\prime}) &=& \int \frac{ d^{4}k}{(2\pi)^{4}} [\exp{(ik_{\mu} u_{\mu})}]{g}(k) \nonumber \\
&\times&[\exp{(ik_{\nu} u_{\nu}^{\prime})}]^{*},
\label{Glivre}
\end{eqnarray}
being $g(k)$ the propagator in the momenta space. It is found through application $(\partial_{\mu}\partial_{\mu} - m^{2})$ at Eq.~(\ref{Glivre}). After this, we use Eq.~(\ref{GlivreEqDif}) and take into account that 
\begin{eqnarray}
\delta^{4}(u-u^{\prime}) = \int \frac{ d^{4}k}{(2\pi)^{4}} [\exp{(ik_{\mu} u_{\mu})}][\exp{(ik_{\nu} u_{\nu}^{\prime})}]^{*}.
\label{ondas}	
\end{eqnarray}
The result is 
\begin{eqnarray}
{g}(k) = \frac{1}{k^{2}+m^{2}},
\label{gemM}
\end{eqnarray}
where $k^{2} = k^{2}_{\tau}+k^{2}_{x}+k^{2}_{y}+k^{2}_{z}$. Now, we can go back in Eq.~(\ref{Glivre}) and replace $g(k)$ by Eq.~(\ref{gemM}).

\subsection{Scalar field  propagator under an electromagnetic external field}

Now let us calculate the bosonic propagator under an electric external field, ${\bf{E}} = E_{0}\hat{z}$ and a magnetic external field, ${\bf{B}} = B_{0}\hat{z}$. For this, we will use the minimal coupling in the Euclidean space~\cite{lawrie1}: $\partial_{\mu} \rightarrow \partial_{\mu}+ieA_{\mu}^{ext}$. This is the covariant derivative in the Euclidean space, $D_{\mu}$. We choose the gauge $A_{\mu}^{ext} = (-zE_{0},0,xB_{0},0)$. The propagator, in this case, satisfy the equation
\begin{eqnarray}
(D_{\mu}D_{\mu} - m^{2})G(u,u^{\prime},A) = -\delta^{4}(u-u^{\prime}).
\label{eqdeGreenBosons}
\end{eqnarray}
Notice that $[D_{\mu}^{2},\partial_{\nu}] \neq 0$. Therefore, the planes waves are not eigenfuntions of $D_{\mu}^{2}$ operator.

Let $E_{p}(u)$ the eigenfunctions of operator $D_{\mu}^{2}$. If we get the complete set formed by $E_{p}(u)$, then we will can write
\begin{eqnarray}
G(u,u^{\prime},A) = \int dp E_{p}(u){\mathcal{G}}(p){E}^{*}_{p}(u^{\prime}).
\label{Gmag1}
\end{eqnarray}
To find ${\mathcal{G}}(p)$, we have to apply the operator  $\left(D_{\mu}^{2} - m^{2}\right)$ over Eq.~(\ref{Gmag1}), to use the Eq.~(\ref{eqdeGreenBosons}) and write the delta function in terms of $E_{p}(u)$. 

Therefore, let us solve the eigenvalue equation
\begin{eqnarray}
D_{\mu}^{2}E_{p}=p^2E_{p}.
\label{eqautovalorB}
\end{eqnarray}
According to our gauge, the operator  $D_{\mu}^{2}$ becomes
\begin{eqnarray}
D_{\mu}^{2} &=& \partial_{\tau}^{2}+\partial_{x}^{2}+\partial_{y}^{2}+\partial_{z}^{2} - 2i\omega_{E}z\partial_{\tau}  \nonumber \\ 
&&+2i\omega_{B}x\partial_{y} - \omega^{2}_{E} z^2 - \omega^{2}_{B} x^2.
\label{Pidois}
\end{eqnarray}
being $\omega_{E} \equiv e E_{0}$ and $\omega_{B} \equiv e B_{0}$.

From Eq.~(\ref{Pidois}), we notice that variables $(\tau,z)$ and $(x,y)$ are coupling. Thus, to solve Eq.~(\ref{eqautovalorB}), we make the ansatz \cite{lawrie1,Emerson,EmersonRBEF},
\begin{eqnarray}
{E}_{p}(u) = C  \exp\left[i(\omega_{E} a_{\tau} \tau + \omega_{B} a_{y} y )\right]X(x)Z(z),
\label{ansatzeB}
\end{eqnarray}
where $C$ is a normalization constant.

Replacing the ansatz on the equation (\ref{eqautovalorB}), we get   
\begin{eqnarray}
\left(\frac{Z''(z)}{Z} - \omega^{2}_{E}(z-a_{\tau})^{2}\right) &=& \nonumber \\ 
- \left(\frac{X''(x)}{X} - \omega^{2}_{B}(x+a_{y})^{2}-p^{2}\right) &\equiv& - const.  
\label{eq.HermiteB}
\end{eqnarray}
To obtain finite solutions in $z$ variable, we choose $const = \omega_{E}(2n+1)$, $n = 0,1,2, \cdots$. In $x$ variable, the choose that produces finite solutions is $-p^{2}-const = \omega_{B}(2\ell+1)$, $\ell = 0,1,2, \cdots$. Thus,
\begin{eqnarray}
p^{2} = -\left[\omega_{E}(2 n +1)+\omega_{B}(2 \ell +1)\right], \, n,\ell = 0,1,2,\cdots.
\label{p 2B}
\end{eqnarray}

The Eq.~(\ref{p 2B}) gives the finite solutions in terms of Hermite polynomials, namely
\begin{eqnarray}
	X_{\ell}(x) &=&\frac{1}{\sqrt{2^{\ell} \ell !}%
	}\left( \frac{\omega_{B} }{\pi }\right) ^{\frac{1}{4}} \exp\left[{-\omega_{B}(x+a_{y})^{2}/2}\right] \nonumber \\
	&\times&H_{\ell }\left[ \sqrt{\omega_{B} }(x+a_{y})\right],
	\label{solucaoX}
\end{eqnarray}
and
\begin{eqnarray}
	Z_{n}(z) &=&\frac{1}{\sqrt{2^{n} n !}%
	}\left( \frac{\omega_{E} }{\pi }\right) ^{\frac{1}{4}} \exp\left[{-\omega_{E}(z-a_{\tau})^{2}/2}\right] \nonumber \\
	&\times&H_{n }\left[ \sqrt{\omega_{E} }(z-a_{\tau})\right].
	\label{solucaoZ}
\end{eqnarray}
 The Eqs.~(\ref{solucaoX}) and (\ref{solucaoZ}) can be written in terms of the Hermite functions~\cite{alemao}
\begin{eqnarray*}
	h_{m}(s)\equiv\frac{1}{\sqrt{2^{m }m !\sqrt{\pi}}}\exp\left(-\frac{{s}^{2}}{2}\right) H_{m }(s).
\end{eqnarray*}

These functions form a complete set, i.e.,
\begin{eqnarray}
\int_{-\infty}^{+\infty}ds \ h_{\ell}(s) \ h_{n}(s) = \delta_{\ell,n}
\label{ortoB}
\end{eqnarray}
and
\begin{eqnarray}
\sum_{n = 0}^{+\infty} \ h_{n}(s) \ h_{n}(s^{\prime}) = \delta(s-s^{\prime}).
\label{completezahB}
\end{eqnarray}

Therefore, we get
\begin{eqnarray}
{E}_{p}(u) &=& (\omega_{E}\omega_{B})^{\frac{1}{4}}\exp\left[i(\omega_{E}a_{\tau}\tau + \omega_{B}a_{y}y)\right] \nonumber \\
&\times&h_{\ell}[\sqrt{\omega_{B}}(x+a_{y})]h_{n}[\sqrt{\omega_{E}}(z-a_{\tau})].
\label{autofuncaoB}
\end{eqnarray}

From Eqs.~(\ref{autofuncaoB}) and (\ref{completezahB}) we can show the completeness relation
\begin{eqnarray}
\int d{p} {E}_{p}(u)E^{*}_{p}(u^{\prime}) &=& \sum_{n = 0}^{+\infty}\sum_{\ell = 0}^{+\infty}\int\frac{\omega_{E}\omega_{B}}{(2\pi)^{2}}d{a}_{\tau} d{a}_{y}{E}_{p}(u)E^{*}_{p}(u^{\prime})  \nonumber \\
&=& \delta^{4}(u-u^{\prime}).
\label{completezaB}
\end{eqnarray}
and the orthogonality
\begin{eqnarray}
\int d^{4}u{{E}}_{p}(u)  E^{*}_{p^{\prime}}(u)&=& (2\pi)^{2}\delta[\omega_{E}(a_{\tau}-a_{\tau}^{\prime})]\delta[\omega_{B}(a_{y}-a_{y}^{\prime})] \nonumber \\ 
&\times &\delta_{n,n^{\prime}}\delta_{\ell,\ell^{\prime}}.
\label{ortogonalidadeB}
\end{eqnarray}

Thus, the scalar field propagator under external electromagnetic field is given in terms of the Ritus' eigenfunctions, $E_{p}(u)$:
\begin{eqnarray}
G(u,u^{\prime},A) &=& \frac{\omega_{E}\omega_{B}}{(2\pi)^{2}}\sum_{n=0}^{+\infty}\sum_{\ell=0}^{+\infty}\int d{a}_{\tau} d{a}_{y} E_{p}(u){\mathcal{G}}(p){E}^{*}_{p}(u^{\prime}), \nonumber \\ 
\label{Gmag}
\end{eqnarray}
with $E_{p}(u)$ given by Eq.~(\ref{autofuncaoB}) and $p^2$ written in Eq.~(\ref{p 2B}).

Applying $(D_{\mu}^{2}-m^{2})$ at relation given in the Eq.~(\ref{Gmag}) and using the Eqs.~(\ref{eqdeGreenBosons}), (\ref{eqautovalorB}), (\ref{p 2B}) and (\ref{completezaB}), we found ${\mathcal{G}}(p)$ in the Euclidean space
\begin{eqnarray}
{\mathcal{G}}(p)=\frac{1}{\omega_{E}(2 n+1)+\omega_{B}(2 \ell+1)+m^{2}}.
\end{eqnarray}
Note that ${\mathcal{G}}(p)$ actually does not depends of $p$, but only of eletric and magnetic fields and the Landau levels, $(n,\ell)$.

\subsection{Free spinor field propagator}

Let us find the well know free Dirac field propagator by the Ritus' method. In the Euclidean space, we have
\begin{eqnarray*}
({\slashed{\partial}}+im)\psi(u)=0,
\end{eqnarray*}
where
\begin{eqnarray*}
{\slashed{\partial}} &=&\gamma_{\mu}\partial_{\mu} \ ; \ \{\gamma_{\mu},\gamma_{\nu}\} =-2\mathrm{\delta}_{\mu\nu}.
\end{eqnarray*}
Explicitly
\begin{equation}
\mathbb{\gamma}_{\mathrm{0}} =-i\left( 
\begin{array}{cc}
\mathrm{0}_{2} & \mathbb{I}_{2} \\ 
\mathbb{I}_{2} & \mathrm{0}_{2}%
\end{array}\right) \,;\,
\mathbb{\gamma}_{\mathrm{j}} =\left( 
\begin{array}{cc}
0_{2} & -\mathbb{\sigma}_{\mathrm{j}} \\ 
\mathbb{\sigma}_{\mathrm{j}} & 0_{2}%
\end{array}\right),
\label{matrizes de Dirac}
\end{equation}
being $\mathbb{\sigma}_{1} = \mathbb{\sigma}_{x}\,,\mathbb{\sigma}_{2} = \mathbb{\sigma}_{y}$ and $\mathbb{\sigma}_{3} = \mathbb{\sigma}_{z}$ the Pauli matrices.

The free propagator satisfy the equation 
\begin{eqnarray}
({\slashed{\partial}}+im)S(u-u^{\prime}) = \delta^{4}(u-u^{\prime}).
\label{greenlivre}
\end{eqnarray}

Since $[{\slashed{\partial}},{\partial}_{\nu}] = 0$, eigenfunctions of operator ${\partial}_{\nu}$, namely, planes waves, are also eigenfunctions of ${\slashed{\partial}}$, with $\slashed{k}$ eigenvalue:
\begin{eqnarray}
\slashed{\partial}[\exp{(i{k}_{\mu} u_{\mu})}] = \left[\exp{(i{k}_{\mu} u_{\mu})}\right]i\slashed{k},
\label{ondaplanaF}
\end{eqnarray}
in this order.

Thus, expanding the Dirac propagator in terms of planes waves, we obtain 
\begin{eqnarray}
S(u-u^{\prime}) &=& \int \frac{d^{4}k}{(2\pi)^{4}}[\exp{(i{k}_{\mu} u_{\mu})}]\tilde{s}(k) \nonumber \\
&\times&[\exp{(i{k}_{\nu} u^{\prime}_{\nu})}]^{*}.
\label{Slivre}
\end{eqnarray}

The free propagator in the momenta space, $\tilde{s}(k)$, can be found by application of operator $({\slashed{\partial}}+im)$ at equation (\ref{Slivre}), and by use of the Eqs.~(\ref{ondaplanaF}), (\ref{greenlivre}) and Eq.~(\ref{ondas}), namely 
\begin{eqnarray}
\tilde{s}(k) =  \frac{i(\slashed{k}-m)}{k^{2}+m^{2}}.
\label{propemM}
\end{eqnarray}

\subsection{Spinor field propagator under an electromagnetic external field}

Take into account the minimal coupling, the Dirac equation under both electric and magnetic fields in the $z$ direction is given by
\begin{eqnarray*}
({\slashed{D}} +i m )\Psi(u) = 0,
\label{Eq. Dirac}
\end{eqnarray*}
where again $D_{\mu} = \partial_{\mu}+ie{A}_{\mu}^{ext}$ and $A_{\mu}^{ext} = (-zE_{0},0,xB_{0},0)$. The Feynman propagator is 
\begin{eqnarray}
({\slashed{D}} +i m )S(u,u^{\prime},A) =  \delta^{4}(u-u^{\prime}).
\label{eq. de Green}
\end{eqnarray}
Note that $[\slashed{D},{\partial}_{\nu}]=ie\gamma_{\mu}[A_{\mu}^{ext},{\partial}_{\nu}]\neq 0$. Therefore, the wave planes are not eigenfunctions of operator $\slashed{D}$. However, $[{\slashed{D}}^{2},\slashed{D}] = 0$. Thus, let us find the eigenfunctions of quadratic Dirac operator.

If we could find the eigenfunctions $\mathbb{E}_{p}$ of operator $\slashed{D}$, or equivalently $\slashed{D}^2$, we can write the fermionic propagator by expansion in terms of them. 

We can show that
\begin{eqnarray*}
	{\slashed{D}}^{2} = \left(i e [\gamma_{\mu},\gamma_{\nu}]F_{\mu\nu}-4 D_{\mu}^{2}\right)/4,
\end{eqnarray*}
where $D_{\mu}^{2}$, is given by Eq.~(\ref{Pidois}) and the only no null values of $F_{\mu\nu}$ is $F_{03} = -F_{30} = E_{0}$ and $F_{12} = - F_{21} = B_{0}$. Using explicitly the $\mathrm{\gamma}$ matrices write in Eq.~(\ref{matrizes de Dirac}), we have
\begin{eqnarray}
	\slashed{D}^{2} = \omega_{E}  (\tilde{\mathbb{I}}_{2}\otimes \mathrm{\sigma}_{z})+\omega_{B}  \left(\mathbb{I}_{2}\otimes \mathbb{\sigma}_{z}\right)- D_{\mu}^{2},
\end{eqnarray}
where
\begin{equation}
 \tilde{\mathbb{I}}_{2} \equiv \left( 
\begin{array}{cc}
1 & 0 \\ 
0& -1%
\end{array}\right).
\end{equation}

From bosonic case, we know that the operator $D_{\mu}^{2}$ has its eigenvalues given by $\left[-\omega_{E}(2 n+ 1) -\omega_{B}(2 \ell +1) \right] $ and its eigenfunctions given by $E_{p}(u)$ (see Eq.~(\ref{autofuncaoB})). Therefore, we have just that find the spin related part eigenvalue of $\slashed{D}^{2}$, i.e.,$ [\omega_{E} (\tilde{\mathbb{I}}_{2}\otimes \mathbb{\sigma}_{z})+\omega_{B}  \left(\mathbb{I}_{2}\otimes \mathbb{\sigma}_{z}\right)]$.

We need find the matrices ${\mathbf{\Omega}}_{\sigma_{\mathrm{E}}}$ and  ${\mathbf{\Omega}}_{\sigma_{\rm{B}}}$ such that
\begin{eqnarray}
(\tilde{\mathbb{I}}_{2}\otimes\sigma_{z})\,{\mathbf{\Omega}}_{\mathbb{\sigma}_{\rm{E}}}&=&\mathbb{\sigma}_{\rm{E}}\,{\mathbf{\Omega}}_{\mathbb{\sigma}_{\rm{E}}},  \nonumber \\
\left(\mathbb{I}_{2}\otimes\mathbb{\sigma}_{z}\right){\mathbf{\Omega}}_{\mathbb{\sigma}_{\rm{B}}}&=&\mathbb{\sigma}_{\rm{B}}\,{\mathbf{\Omega}}_{\mathbb{\sigma}_{\rm{B}}},
\label{ansatze5}
\end{eqnarray}
with $\mathbb{\sigma}_{\rm{E}} = \mathrm{\sigma}_{\rm{B}} = \pm 1.$

After we make the tensorials products in Eq.~(\ref{ansatze5}), we easily found the structure of ${\mathbf{\Omega}}_{\mathbb{\sigma}_{E}}$ and ${\mathbf{\Omega}}_{\mathbb{\sigma}_{B}}$. Therefore, the eigenfunctions of ${\mathbb{\slashed{D}}}^{2}$ are given by~\cite{EmersonRBEF,Elizalde1}

\begin{eqnarray}
{\mathbb{E}}_{p}(u) = \sum_{\mathbb{\sigma}_{\rm{E}},\mathbb{\sigma}_{\rm{B}=\pm 1}}E_{p,\mathrm{\sigma_{\rm{E}}},\mathrm{\sigma_{\rm{B}}}}(u)\,{\mathbf{\Omega}}_{\mathbb{\sigma}_{\mathrm{E}},\mathrm{\sigma}_{\mathrm{B}}}\,,
\label{ansatzenova}
\end{eqnarray}
where
\begin{eqnarray}
{\bf{\Omega}}_{\sigma_{\rm{E}},\sigma_{\rm{B}}} &=& {\bf{diag}}\left(\frac{}{}\delta_{1,\sigma_{\rm{E}}}\,\delta_{1,\sigma_{\rm{B}}}\,;\,\delta_{-1,\sigma_{\rm{E}}}\,\delta_{-1,\sigma_{\rm{B}}}\,; \right. \nonumber \\ &&\left.\delta_{-1,\sigma_{\rm{E}}}\,\delta_{1,\sigma_{\rm{B}}}\,;\,\delta_{1,\sigma_{\rm{E}}}\,\delta_{-1,\sigma_{\rm{B}}}\frac{}{}\right),
\label{diag}
\end{eqnarray}
and $E_{p,\sigma_{\rm{E}},\sigma_{\rm{B}}}(u)$ are given by Eq.~(\ref{autofuncaoB}). Thus,
\begin{eqnarray}
\slashed{D}^{2} {\mathbb{E}}_{p} &=& \left[\omega_{E}(2 n+1 + \sigma_{E})+\omega_{B}(2 \ell+1 + \sigma_{B})\right]{\mathbb{E}}_{p}. \nonumber \\ 
&& 
\label{eq.autovalor}
\end{eqnarray}

Although we already know the eigenfunctions of operator $\slashed{D}$, i.e., the Ritus' eigenfunctions ${\mathbb{E}}_{p}(u)$ given by Eq.~(\ref{ansatzenova}), we do not know what are its eigenvalues, and for this reason, we still cannot expand the fermionic propagator as we did in Eqs.~(\ref{Glivre}), (\ref{Gmag}) and (\ref{Slivre}) in terms of $\mathbb{E}_p$. To solve this problem, Ritus, postulated the relation
\begin{eqnarray}
\slashed{D} \, \mathbb{E}_{p} = \mathbb{E}_{p}\, i \slashed{\bar{p}},
\label{pbarra}
\end{eqnarray}
in this order, for a specific four-vector $\bar{p}$.

In the {\bf{appendix A}}, we calculate explicitly the four-vector $\bar{p}$. The result is
\begin{eqnarray}
\bar{p}_{\mu} &=&\left(\sqrt{\omega_{E}(2 n+1 + \sigma_{\rm{E}})},0,\sqrt{\omega_{B}(2 \ell+1 + \sigma_{\rm{B}})},0\right), \nonumber \\
& & \, n,\ell = 0,1,2,\cdots.
\label{componentes}
\end{eqnarray}

Take into account the Eq.~(\ref{completezaB}) and that, 
\begin{eqnarray*}
\sum_{\mathbb{\sigma}_{\rm{E}},\mathbb{\sigma}_{\rm{B}=\pm 1}} \,\,  \sum_{\mathbb{\sigma}_{\rm{E^{\prime}}},\mathbb{\sigma}_{\rm{B^{\prime}}= \pm 1}} {\mathbf{\Omega}}_{\mathbb{\sigma}_{\mathrm{E}},\mathrm{\sigma}_{\mathrm{B}}} \,\, {\mathbf{\Omega}}_{\mathbb{\sigma}_{\mathrm{E^{\prime}}},\mathrm{\sigma}_{\mathrm{B^{\prime}}}} = \mathbb{I}_4,
\end{eqnarray*}
is easy show that
\begin{eqnarray}
\int d{p} \, \mathbb{E}_{p}(u) \, \mathbb{E}_{p}^{\dagger}(u^{\prime}) = \delta^{4}(u - u^{\prime}),
\label{CompletezaBB}
\end{eqnarray}
and 
\begin{eqnarray}
\int d^{4}u \, \mathbb{E}_{p}(u) \, \mathbb{E}_{p^{\prime}}^{\dagger}(u) &=& (2\pi)^{2}\delta[\omega_{E}(a_{\tau}-a_{\tau}^{\prime})]\delta[\omega_{B}(a_{y}-a_{y}^{\prime})] \nonumber \\ 
&\times &\delta_{n,n^{\prime}}\delta_{\ell,\ell^{\prime}}.
\label{ortogonalidadeBB}
\end{eqnarray}

Now, we finally can expand the fermionic field propagator in terms $\mathbb{E}_{p}$, namely 
\begin{eqnarray}
S(u,u^{\prime},A) = \int \,dp\, \mathbb{E}_{p}(u){\mathcal{S}}(p){{\mathbb{E}}^{\dagger}}_{p}(u^{\prime}),
\label{Smag1}
\end{eqnarray}

Replacing the expansion given in Eq.~{(\ref{Smag1})} within Eq.~(\ref{eq. de Green}) and using the Eqs.~({\ref{pbarra}}) and (\ref{CompletezaBB}), we obtain 
\begin{eqnarray}
{\mathcal{S}}(\bar{p}) = \frac{i(\slashed{\bar{p}}-m)}{\bar{p}^{2}+m^2},
\end{eqnarray}
with the $\bar{p}_{\mu}$ components given by Eq.~(\ref{componentes}). We also have used that $\slashed{\bar{p}}^{2} = - \bar{p}^{2}$.
Finally, the Dirac propagator under an external constant electromagnetic field is 
\begin{eqnarray}
S(u,u^{\prime},A) &=& \frac{\omega_{E}\omega_{B}}{(2\pi)^{2}}\sum_{n=0}^{+\infty}\sum_{\ell=0}^{+\infty}\int d{a}_{\tau} d{a}_{y} \mathbb{E}_{p}(u){\mathcal{S}}(\bar{p}){{\mathbb{E}}^{\dagger}}_{p}(u^{\prime}),\nonumber\\
&&
\label{Smag}
\end{eqnarray}
with $\mathbb{E}_p(u)$ write in Eq.~(\ref{ansatzenova}).
\vspace{0.25 cm}
\section{Conclusions / Comments}

In this paper, we rescue the Ritus' idea to calculate the Feynman propagator under magnetic and electric external fields. First, we used the Klein Gordon equation and the minimal coupling prescription to find the Ritus' eigenfunctions and its eigenvalues. After that, we apply the method at the Dirac equation. In this case, we have had an additional problem: we did not know the eigenvalues associated with the Dirac operator inside the constant electromagnetic field. Thus, one more condition was necessary to find the fermionic propagator. We define that $\slashed{D} \, \mathbb{E}_{p} = \mathbb{E}_{p}\, i \slashed{\bar{p}}$ for some $\bar{p}_{\mu}$ to be found. In the appendix $A$, we calculate explicitly the components of $\bar{p}$. The Ritus' method is a sensational tool to find the Feynman propagator in an external field. As the bosonic propagator, as well as the fermionic propagator were written in a diagonal form, the expressions involved became simpler than those
obtained by other methods, for example, the proper time method. With this manuscript, we expect that the Ritus' method become more widespread in the wider academic community.

\section*{Acknowledgments}

This paper is dedicated to the memory of Jos\'e Roberto Ferreira de Sarges. E.B.S.C thanks to PROPIT/UNIFESSPA for financial support through project PIBIC0663016710123. 

\section*{Appendix A: The components of $\bar{p}_{\mu}$}

Now let us demonstrate the $\bar{p}_{\mu}$ components written in ~(\ref{componentes}). For this, we will write explicitly the left side of Eq.~(\ref{pbarra}). Take into account the Eq.~(\ref{ansatzenova}), we have
\begin{equation}
{\mathbb{E}}_{p} =\left( 
\begin{array}{cccc}
E_{1,1} & 0 & 0 & 0 \\ 
0 & E_{-1,-1}  & 0 & 0  \\%
0 & 0 & E_{-1,1} & 0 \\
0 & 0 & 0 & E_{1,-1}  
\end{array}\right).
\label{matriz E} 
\end{equation}
Using the $\rm{\gamma}$ matrices given in Eq.~(\ref{matrizes de Dirac}) we get 
\begin{widetext}
	\begin{eqnarray*}
		\slashed{D} \, {\mathbb{E}}_{p} =
	\end{eqnarray*}
	\begin{equation}
	\left( 
	\begin{array}{cccc}
	0                 &           0                     & (-i D_{0}-D_{3})E_{-1,1}                 & (-D_{1}+i D_{2})E_{1,-1} \\ 
	0                 &0                                & (- D_{1}-i{D}_{2})E_{-1,1} &      (-i{D}_{0}+ D_{3})E_{1,-1}            \\
(-i{D}_{0}+ D_{3})E_{1,1}& (D_{1}-i D_{2})E_{-1,-1}& 0                         &    0                                      \\
(D_{1}+ i{D}_{2})E_{1,1}    &  (-i{D}_{0}- D_{3})E_{-1,-1}          & 0             & 0 
	\end{array}\right),
	\label{matriz gammaPiE} 
	\end{equation}
\end{widetext}
where $D_{0} = \partial_{\tau} - i \omega_{E} z$, $D_{1} = \partial_{x}$, $D_{2} = \partial_{y} + i \omega_{B} x$ and $D_{3} = \partial_{z}$.

The right side of Eq.~(\ref{pbarra}) is 
\begin{widetext}
	\begin{eqnarray*}
	{\mathbb{E}}_{p} \,i \slashed{\bar{p}}=
	\end{eqnarray*}
	\begin{equation}
	\left( 
	\begin{array}{cccc}
	0                 &           0                     & (\bar{p}_{0}-i\bar{p}_{3})E_{1,1}                 &  (-i\bar{p}_{1}-\bar{p}_{2})E_{1,1} \\ 
	0                 &0                                &(-i\bar{p}_{1}+i\bar{p}_{2})E_{-1,-1}      &      (\bar{p}_{0}+i\bar{p}_{3})E_{-1,-1}                \\
(\bar{p}_{0}+i\bar{p}_{3})E_{-1,1}     & (i\bar{p}_{1}+\bar{p}_{2})E_{-1,1}& 0                         &    0                                      \\
	(i\bar{p}_{1}-\bar{p}_{2})E_{1,-1}  &  	(\bar{p}_{0}-i\bar{p}_{3})E_{1,-1}          & 0             & 0 
	\end{array}\right).
	\label{matriz pbarra} 
	\end{equation}
\end{widetext}

After we compare the Eqs.~(\ref{matriz gammaPiE}) and (\ref{matriz pbarra}) and solve four sets of differential coupling equations, we find that the components of $\bar{p}_{\mu}$ satisfy
\begin{eqnarray*}
\bar{p}_{0}^{2} + \bar{p}_{3}^{2} &=& \omega_{E}(2 n +1 + \sigma_{\rm{E}}),
\end{eqnarray*}
and
\begin{eqnarray*}
\bar{p}_{1}^{2} + \bar{p}_{2}^{2} &=& \omega_{B}(2 \ell +1 + \sigma_{\rm{B}}),	
\end{eqnarray*}
for $n, \ell = 0,1,2, \cdots$ and $\sigma_{\rm{E}} = \sigma_{\rm{B}} = \pm 1$. To an appropriate coordinate system, we can fix $\bar{p}_{1} = \bar{p}_{3} = 0$, and thus, the Eq.~(\ref{componentes}) is demonstrated.

\end{document}